\documentclass{eas}
\usepackage[dvips]{graphicx}

\begin{document}

\title{Soliton solutions in relativistic field theories and
gravitation}

\author{Joaquin Diaz-Alonso}
\address{LUTH, Observatoire de Paris, CNRS, Universit\'e Paris
Diderot. 5 Place Jules Janssen, 92190 Meudon, France}
\secondaddress{Departamento de Fisica, Universidad de Oviedo.
Avda. Calvo Sotelo 18, E-33007 Oviedo, Asturias, Spain}
\author{Diego Rubiera-Garcia}
\sameaddress{2}

\begin{abstract}

We report on some recent results on a class of relativistic
lagrangian field theories supporting non-topological soliton
solutions and their applications in the contexts of Gravitation
and Cosmology. We analyze one and many-components scalar fields
and gauge fields.

\end{abstract}

\maketitle

\section{Introduction}

The well-known Born-Infeld (BI) model (Born \& Infeld
\cite{borninfeld1934}) was historically proposed to remove the
divergence of the electron's self-energy in classical
electrodynamics. Aside from electrostatic soliton solutions in
three space dimensions, the model exhibits dyon and bidyon-like
solutions (Chernitskii \cite{chern1999}), electric-magnetic
duality (Gibbons \& Rasheed \cite{gibbons1995}) and special
properties of wave propagation, belonging to the class of
``completely exceptional" theories (Boillat \cite{boillat1970}).

Recently there has been a renewed interest in BI theory
and its non-abelian extensions since they appear at different
levels of string theory (Gibbons \cite{gibbons1998}). However,
there are many other physical contexts where BI-like models have
been used, such as the description of dark energy in Cosmology (e.g.
F\"uzfa \& Alimi \cite{fuzfa2006a}) and the phenomenological
description of the nucleon structure (Deser et al
\cite{deser1976}; Pavlovski \cite{pavlovsky2002}).

On the other hand, many studies on generalized electromagnetic and
gauge field theories coupled with gravitation have been devoted to
the analysis of particle-like solutions. Although several theorems of the
70's (Coleman \cite{coleman1977}) forbid the existence of static,
finite-energy solutions of the pure Yang-Mills theory, such
solutions were found in the Einstein-Yang-Mills system (Bartnick
\& McKinnon \cite{bartnick1988}). Later it was shown that similar
glueball solutions also exist in BI-like models in flat space
(Gal'tsov \& Kerner \cite{galtsov2000}) as well as in curved space
(e.g. Wirschins \emph{et al} \cite{wirschins2001}) (see Volkov \&
Gal'tsov (\cite{volkov1999}) for a review and references on
non-abelian solitons).

\section{The models}

The BI generalization of classical electrodynamics is by no means
unique and, from the point of view of soliton solutions, BI theory
is only a particular example of a large class, which has been
exhaustively determined in Diaz-Alonso \& Rubiera-Garcia
(\cite{dr2007b}). Here we deal with the generalized gauge field
problem, by considering gauge-invariant lagrangians which are
given functions $L=\varphi(X,Y)$ of the two quadratic field
invariants $X=-\frac{1}{2}\sum_aF_{\mu\nu}^aF^{\mu\nu a}
\hspace{2pt};\hspace{2pt} Y = -\frac{1}{2}
\sum_aF_{\mu\mu}^aF^{*\mu\nu a}$, defined in a domain ($\Omega
\subseteq \Re^{2}$) which is assumed to be open and connected and
including the vacuum ($X=Y=0$). In calculating the field
invariants we use the ordinary definition of the trace, although
other definitions are possible (Tseytlin \cite{tseytlin1997}). We
also require the condition $\varphi(X,Y)=\varphi(X,-Y)$ to be
satisfied, in order to preserve parity invariance. Moreover we
restrict our analysis to ``physically admissible theories", which
we define by the requirement of the vanishing of the vacuum energy
($\varphi(0,0)=0$), as well as the positive definiteness of the
energy, which demands the minimal \emph{necessary and sufficient}
condition

\begin{eqnarray}\label{1}
\rho^s\geq \left(X+\sqrt{X^2+Y^2}\right)\frac{\partial
\varphi}{\partial X}+ Y\frac{\partial \varphi}{\partial
Y}-\varphi(X,Y)\geq 0\hspace{2pt},\hspace{2pt}
\end{eqnarray}
to be satisfied in the entire domain of definition ($\Omega$). The field equations read

\begin{equation}\label{2}
    \sum_bD_{ab\mu}\left[\frac{\partial \varphi}{\partial X}F^{\mu\nu b}+
\frac{\partial \varphi}{\partial Y}F^{*\mu\nu b}\right]=0\hspace{2pt},\hspace{2pt}
\end{equation}
where $D_{ab\mu}\equiv\delta_{ab}\partial_{\mu}
+g\sum_cC_{abc}A_{c\mu}$. We next consider the electrostatic
spherically symmetric solutions (ESS) ($\vec{E}_a(r)=
-\vec{\nabla}(\phi_a(r)) = -\phi'_a(r) \frac{\vec{r}}{r}
\hspace{1pt}; \hspace{1pt}\vec{H}_a=0$). When this substitution is
done in (\ref{2}) we get two sets of equations

\begin{eqnarray}\label{3}
\nu=0 \rightarrow \vec{\nabla}\left(\frac{\partial \varphi}{\partial X}\vec{\nabla}\phi_a(r)\right)=0
\hspace{4pt};\hspace{4pt}
\nu=i=1,2,3 \rightarrow \sum_{bc}C_{abc}\phi_b(r)\phi'_c(r)=0\hspace{2pt}.\hspace{2pt}
\end{eqnarray}
The first set leads to the first integrals $r^2\frac{\partial
\varphi}{\partial X}\phi'(r) = Q_a$ ($Q_a$ being integration
constants, identified as \emph{color charges}) which coincide with
the set of first integrals of the field equations for static,
spherically symmetric solutions (SSS) of a multicomponent scalar
field theory with a lagrangian density given by

\begin{equation}\label{4}
L=f(\sum_a\partial_{\mu}\phi_a\partial^{\mu}\phi_a)\equiv f(X) =
-\varphi(-X,Y=0)\hspace{2pt}.\hspace{2pt}
\end{equation}
The solutions of  Eqs.(\ref{3}) for the multiscalar or gauge cases
take the form $\phi_a(r) = \frac{Q_a}{Q}\left(\phi(r,Q) +
\alpha_a\right)$, where $Q=\sqrt{Q_a^2}$ is the mean-square
(\emph{scalar} or \emph{color}) charge and $\alpha_a$ are
integration constants. The function $\phi(r,Q)$ is the solution of
the one-component scalar field theory resulting from the restricted
lagrangian $f(X=\partial_{\mu}\phi\partial^{\mu}\phi)$. In the
gauge case the second set of equations (\ref{3}) restricts the
possible values of the integration constants
$\alpha_a=\frac{Q_a}{Q}\alpha$ and then the ESS solutions take the
form $\phi_a(r)=\frac{Q_a}{Q}\left(\phi(r,Q)+\alpha\right)$ where
now $Q$ is the mean-square \emph{color} charge.

Thus, the form of the ESS solutions of generalized gauge field
theories coincides with that of the SSS solutions of an
associated one-component scalar field theory. The energy of the
ESS solutions, when finite, is twofold the energy of the
associated SSS solutions, which reads

\begin{equation}\label{6}
    \varepsilon(Q)=-4\pi\int_0^{\infty}r^2f\left(-\phi^{'2}(r)\right)dr=
Q^{3/2}\varepsilon(Q=1).
\end{equation}

The requirement of convergence of this integral leads to a
classification of the admissible models according to the central
(\underline{A-cases}) and asymptotic (\underline{B-cases})
behaviors of the soliton field strength (see Fig.1). Cases
\emph{A-1} and \emph{A-2} correspond to infinite (but integrable) and finite
soliton field strengths, respectively. Cases \emph{B-1}, \emph{B-2} and
\emph{B-3} correspond to asymptotic dampings of the soliton field
strength which are slower than coulombian, coulombian, or faster
than coulombian, respectively (Diaz-Alonso \& Rubiera-Garcia \cite{dr2007a}).
\begin{figure}
\begin{center}
\includegraphics[width=6cm,height=4.05cm]{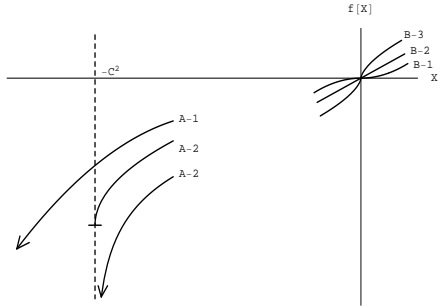}
\caption{Different possible central and asymptotic behaviors of
the admissible models}
\end{center}
\end{figure}

For each admissible scalar model supporting finite-energy SSS
solutions there exists an infinite family of admissible gauge field
theories supporting similar finite-energy ESS solutions, obtained
from (\ref{4}) and the admissibility conditions (\ref{1}).

The linear stability of the SSS and ESS solutions requires the
energy (\ref{6}) to be a minimum against small charge-preserving
perturbations of the fields and potentials in the gauge and scalar
cases. In the latter we found that all finite-energy SSS solutions
of admissible models are always linearly stable while in the
former the following supplementary condition for stability must be
satisfied

\begin{equation}\label{7}
    \frac{\partial \varphi}{\partial X}-2X\frac{\partial^2 \varphi}{\partial Y^2}>0 \hspace{2pt},
\hspace{2pt} \forall(X,Y=0)\hspace{2pt}.\hspace{2pt}
\end{equation}

As a non-trivial example of this class of soliton-supporting
theories we introduce here the family of generalized
electromagnetic field models

\begin{equation}\label{8}
\varphi(X,Y)=X/2+\lambda X^a+\beta Y^2 (a>3/2
\hspace{2pt},\hspace{2pt}\beta>0),
\end{equation}
which is the simplest generalization of a family of scalar models
(defined from (\ref{8}) by setting $\beta=0$) supporting SSS
soliton solutions, which has been analyzed in Diaz-Alonso \&
Rubiera-Garcia (\cite{dr2007c}). This generalized electromagnetic
family fulfills the admissibility conditions (\ref{1}) and
supports ESS solitons. For certain values of the parameters in the
electromagnetic case ($\lambda<0$ , $a=2$) the non-linear term has
the form of the decoupled part of the four-vertex contribution to
the effective lagrangian of quantum electrodynamics.

\section{Conclusions and outlook}

In summary, we have analyzed scalar and generalized gauge field
theories supporting soliton solutions. Aside from BI-like models
which have been widely used during the last few years, other models have
also recently attracted attention in search of self-gravitating
scalar, electromagnetic and gauge field solitons as well as
regular charged solutions (Ayon-Beato \& Garcia \cite{ayonb}). The
class of models considered here can also be useful in
other physical contexts. For instance, in the extension of the
description of dark energy in Cosmology, as a non-canonical scalar
field (Armendariz-Picon \& Lim \cite{armendariz2005}) or as a
gauge field governed by generalized actions (Dyadichev et al
\cite{dyadichev2002}), in the phenomenological description of
nucleon structure, in generalized gauge theories in higher
dimensions, glueballs, etc.

\end{document}